\documentclass[12pt]{article}

\usepackage{amssymb}

\usepackage{amsmath}
\usepackage{amssymb}
\usepackage{graphicx}

\usepackage{color}

\title{\textbf{Metastatistics of Extreme Values and its Application in Hydrology}}
\author{Massimiliano Ignaccolo\\  \textit{\footnotesize Dept. of Earth \& Ocean Sciences, Nicholas School of the Environment, Duke University,}
\\\textit{\footnotesize Durham, North Carolina} \and Marco Marani\\  \textit{\footnotesize Dept. of Earth \& Ocean Sciences, Nicholas School of the Environment, Duke University,}
\\\textit{\footnotesize Durham, North Carolina} \\ \textit{\footnotesize DICEA and International Center for Hydrology, University of Padova, Padova, Italy}
}

\begin{document}
\maketitle


\begin{abstract}
We present a novel statistical treatment, the ``metastatistics of
extreme events'', for calculating the frequency of extreme events.
This approach, which  is of general validity, is the proper statistical 
framework to address the problem of data with statistical 
inhomogeneities. By use of artificial sequences, we show 
that the metastatistics produce the correct predictions while 
the traditional approach based on the generalized extreme value 
distribution does not.  An application of the metastatistics methodology to the case of 
extreme event to rainfall daily precipitation is also presented . 
\end{abstract}

\section{Introduction}\label{sec:intro}
The importance of predicting the frequency of rainfall extremes is paramount in the design of any major hydraulic 
structure for water resources management and flood control. The established statistical tools used in climate analyses 
and in the engineering practice moved away from the initial concept of probable maximum precipitation \cite{hersh65} 
towards an approach which defines design events based on a specified probability of occurrence. The key concept used 
in this setting is the return period, $T_r$, i.e. the average time interval between two exceedances of the magnitude 
of the event considered, es $h(T_r,\tau)$ ($\tau$ being the time scale of interest. We will focus on the illustrative 
case of $\tau=1$ day in the present paper. The estimation of the event magnitude associated with a specified return 
period (the main design specification) is usually obtained by fitting an 'appropriate' extreme value distribution 
(e.g. \cite{koutso04_I,koutso04_II}). These distributions (EV1, EV2, and EV3, or the Generalized Extreme Value, GEV, 
distributions summarizing them, \cite{frechet27,fishtipp28,gnedenko43,gumbel54,jenki55}) have become the common 
reference for extreme value analyses because their form can be derived theoretically by means of an asymptotic theory 
assuming the number of extreme events (i.e. larger than the magnitude of the event of interest) in any given year to be large.

In practice, estimates of extreme rainfall are made by extracting the annual maxima from the series of precipitated 
amounts (for the time scale of interest), and by fitting the annual maxima series with a Gumbel distribution (EV1) 
to extrapolate the rainfall amount, $h(T_r,\tau)$, associated with the fixed return time. The Gumbel distribution 
is typically used in practice because it is the asymptotic distribution for rainfall maxima provided the distribution 
of rainfall (daily in this case) amounts (say, $x$) does not exhibit a slowly decaying tail (i.e. it decays faster 
than $x^{-\alpha},\alpha>1$). The Lognormal \cite{bion76} and Gamma \cite{groi99} distributions have been used to fit 
daily amounts of rainfall. However, the Authors of \cite{wiltuo05} provide a theoretical framework and exhaustive empirical evidence 
that the probability of exceedance, $\Psi(h)$, of daily rainfall amounts ( with $h\geq10$ mm) is well fitted by a stretched 
exponential function, $\Psi(h)\propto\exp(-h/C)^{w}$, $C,w>0$. Using a statistics jargon we say that the distribution of 
daily amounts of rainfall is right-tail equivalent to a Weibull distribution.

The absence of inverse power law tails in the distribution of daily amounts of rainfall has made the Gumbel distribution 
(EV1) the asymptotic distribution adopted by the hydrological community to fit rainfall extremes. However several Authors
report that the Gumbel distribution underestimates the extreme rainfall amounts (e.g. \cite{wilks93,colesetal03,sissonetal06}). 
The inadequacy of the Gumbel distribution according to \cite{koutso04_I} is due to two factors. (A) A slow rate of 
convergence to the asymptotic distribution. This is so because number of days with non null precipitation in one year is 
bound to be smaller than 365 while the extreme asymptotic theory is valid in the limit when maxima are extracted from a 
large, ideally infinite, number of samples. This problem particularly affects maxima extracted from a Weibull distribution: 
see \cite{harris04,cooketal04_I} for a detail discussion on the rate of convergence.(B) Inhomogeneity of the precipitation 
time series. For the extreme value theorem to apply is necessary the stability of the distribution of the variates from 
which the maxima are extracted. This may not be the case in many practical application: e.g. the functional form of the 
distribution may be fixed but the parameters describing the distribution may be themselves a stochastic variable.      

In\cite{koutso04_I,koutso04_II} Koutsoyannis argues that when the above mentioned factors, (A) and (B), play a role, rainfall extremes 
should be fitted to the Frechet (EV2) distribution (\cite{frechet27}), which has a slower decaying tail than the Gumbel 
distribution. In this manuscript we argue against this choice. A slow rate of convergence to the asymptotic distribution, 
factor (A), \textit{does not} justify the use of the asymptotic form relative to another basin of attraction (from EV1 
to EV2). Instead one should use the ``penultimate'' approximation, the approximation prior to the ``ultimate'' asymptotic 
expression, to fit the maxima series. In the case of variates of the exponential type an analytical expression for the 
penultimate approximation can be derived \cite{harris04,cooketal04_I}. A variate is said to be of the exponential type 
if $\Psi(h)$$=$$\exp{-g(h)}$, where $\Psi(h)$ is the probability of exceeding the threshold $h$, and $g(h)$ is a positive 
function which increases monotonically faster than $\ln(h)$. This is the case of Weibull variates so that we can apply it 
to extreme daily amounts of rainfall. The adoption the EV2 asymptote lacks justification also in the case of inhomogeneity, 
factor (B). To address this problem we need to operate in a fashion similar to the case of inhomogeneous Poisson processes. 
As the rate itself of the Poisson process is a stochastic variable, the calculation of any variables of interest includes 
an integration over all possible values of the rate. The same procedure needs to be adopted in the case of inhomogeneous 
extreme events. We dub this approach as the \textit{Metastatistics} Extreme Value (MEV) one. The MEV approach applied to 
the penultimate approximation is proper tool to examine the occurrences of extremes in series of daily amounts of 
precipitation. We accomplish this by use generated sequences of 50 maxima generated from mixtures (variable 
scale and shape parameter) of Weibull variates. These sequence are used to evaluate the intensity of daily precipitation 
relative to return times up to 1,000 years. The MEV approach yields the correct results while the one based on the EV2 
asymptote results in a systematic overestimation.Moreover we apply the MEV approach to the historic (from 1725 to 2006 
albeit not continuously) time series of daily precipitation amounts collected at Padova, Italy.   

The manuscript is organized as follows. In Section \ref{sec:data} we describe the data set used for this analysis. In 
Section \ref{sec:methods} we briefly summarize the classical extreme event theory, the precondition practice, and 
introduce the metastatistics extreme value (MEV) formula, Our results are exposed in Section \ref{sec:results}, and 
our conclusion are drawn in Section \ref{sec:conclu}.

\section{Data}\label{sec:data}
We consider the daily rainfall amount observed at Padova (Italy) over a span of almost three centuries. During this 
period different, albeit structurally similar, instruments have been adopted at three different locations, which all 
fall within a 1 Km circle. The dataset is freely downloadable, and we refer the reader to \cite{camuffo84} for previous analyses of the Padova time series.

Our data set is composed of three intervals of continuous observations, 1725-1764, 1768-1814, and 1841-2006, which 
are later further divided into five subintervals (1725-1764, 1768-1807, 1841-1880, 1887-2006, and 1841-1920) to 
explore different inhomogeneity hypotheses.
    
\section{Methods}\label{sec:methods}
It is first useful to briefly summarize the Extreme Event Theory, as typically used in hydrology, then present the 
practice of ``penultimate'' approximation for  Weibull variates, and to introduce the use of a metastatics to estimate 
extreme events associated with an assigned return period. 

\subsection{Extreme Value theorem}\label{sec:EVT}
Let $X$ be a stochastic variable and $p(x)$ its probability density function, $F(x)=P(X \leq x)$ its distribution function, 
and $\Psi(x)=1-F(x)$ its complementary distribution function. We can define a new stochastic variable, $Y_{n}$, as the maximum of 
$n$ (an integer number) realizations of the stochastic variable $X$: $Y_{n}=max\{x_{1},x_{2},...,x_{n}\}$. 
$Y_{n}$ is the $n$-sample maximum ($n$ is the cardinality, order, of the maximum) of the ``parent'' stochastic variable $X$. 
If the events generating the realizations of $X$ are independent, the cumulative distribution, $\zeta(y)$, of $Y_{n}$ may 
be expressed as
\begin{equation}
\zeta(y)=[F(y)]^{n}.\label{eq:ev0}
\end{equation}
Upon definition of a renormalized variable $S_{n}=(Y_{n}-b_{n})/a_{n}$, the extreme value theorem 
\cite{frechet27, fishtipp28,gnedenko43} establishes that  
\begin{equation}
\underset{n\rightarrow\infty}{{\lim}} P(S_n<s)= \underset{n\rightarrow\infty}{{\lim}} \zeta(s) = 
\underset{n\rightarrow\infty}{{\lim}} [F(a_{n} \cdot s +b_{n})]^{n}=H(s)\label{eq:ev1}
\end{equation}
where $a_{n}>0$ and $b_{n}$ are ``renormalization'' constants. 
The function $H(s)$ in Eq.~(\ref{eq:ev1})  must be one of the three following types (excluding the degenerate 
case, in which all the probability is concentrated in one value of the random variable):
\begin{equation}
\left.
\begin{aligned}
	& \textrm{EV1 or Gumbel: } & H(s)= &\exp(-\exp(-s)) & \forall s & &\\
	& \textrm{EV2 or Frech{\'e}t: } & H(s)= & \exp(-s^{-\alpha})  & s>0 & \;\;\;\; \textrm{and}  & 0\; s<0 \\
 	& \textrm{EV3 or Weibull: } & H(s)= & \exp(-|s|^{\alpha}) &  s<0 & \;\;\;\; \textrm{and} & 1\; s>0
\end{aligned}
\right\}
\end{equation}

The type of limiting distribution is determined by the property of the distribution of the  
parent variable $X$  \cite{frechet27, fishtipp28,gnedenko43}. In particular,
\begin{equation}
\left.
\begin{aligned}
& \textrm{EV3 } & \omega=\sup{F(x)<1}<+\infty & \textrm{ and} \\
& & \underset{k\rightarrow0^{+}}{\lim}\frac{\Psi(\omega-\lambda x)}{\Psi(\omega-\lambda)}=(-x)^{\theta}\;\; & x<0,\theta>0 \\
& \textrm{EV2} & \underset{\lambda\rightarrow\infty}{\lim}\frac{\Psi(\lambda x)}{\Psi(x)}=x^{-\theta}\;\; & x>0,\theta>0 \\
& \textrm{EV1} & \textrm{in all other cases} &
\end{aligned}
\right\}
\end{equation}
The three asymptotic types, EV1-EV3, can be thought of as special cases of a single Generalized
Extreme Value distribution (GEV) \cite{jenki55}:
\begin{equation}
\zeta(s)=H_{GEV}(s)=\exp\Bigg\{-\Bigg(1+k\frac{s-\mu}{\sigma}\Bigg)_{+}^{-1/k}\bigg\}\label{eq:generalGEV}
\end{equation}
where $(.){+}=max(.,0)$, $\mu$ is the location parameter, $\sigma>0$ is the scale parameter, and $k$ is a shape parameter. 
The limit $k=0$ corresponds to the EV1 distribution, $k>0$ to the EV2 distribution (with $\alpha=1/k,$) and $k<0$ to 
the EV3 distribution (with $\alpha=-1/k$). The function $H_{GEV}(s)$ is usually fitted to the cumulative distribution
of non-normalized maxima, so that the location parameter $\mu$ and the scale parameter $\sigma$ are the renormalization
parameters $b_{n}$ and $a_{n}$ respectively. However, it is important to note that the distribution describing the n-sample 
maximum will strictly be a GEV only for 'large enough' values of $n$. How large the value of $n$ needs to be should 
be determined by analyzing the convergence properties based on the observed realizations of $X$.

\subsection{Penultimate approximation for Weibull variates}\label{sec:penwei}
The expected largest value, $U_{n}$, in $n$ realizations of the variable $x$ is the one that is exceeded
with probability 1/$n$:
\begin{equation}
\label{eq:mode}
\Psi(U_{n})=\frac{{1}}{n}\Longleftrightarrow U_{n}=\Psi^{-1}\bigg(\frac{{1}}{n}\bigg)=F^{-1}\bigg(1-\frac{{1}}{n}\bigg).
\end{equation}
Using this result we can write the cumulative probability $\zeta(y)$
for the $n$-sample maximum $Y_{n}$ as 
\begin{equation}
\zeta(y)=[F(y)]^{n}=[1-\Psi(y)]^{n}=\bigg[1-\frac{{\Psi(y)}}{n\Psi(U_{n})}\bigg]^{n},\label{eq:penult1}
\end{equation}
for $y>U_{n}$ the term $\Psi(y)/\Psi(U_{n})<1$ such that, for $n$ large enough we can substitute the Cauchy 
approximation ($(1-x)^\alpha=1-\alpha x=\exp(-\alpha x)$ when $x \ll 1$) in Eq.(\ref{eq:penult1}) to obtain:
\begin{equation}
\label{eq:penult2}
\zeta(y)=\exp\bigg(-\frac{\Psi(y)}{\Psi(U_{n})}\bigg) \Leftrightarrow -\ln(\zeta(y))=\frac{\Psi(y)}{\Psi(U_{n})} 
\end{equation}
Eq. (\ref{eq:penult2}) is referred to as the ``penultimate'' approximation \cite{cooketal04_I,cramerbook}: the approximation 
prior to the ``ultimate'' approximation being given by the extreme value theorem Eq.(\ref{eq:ev1}). The error made in adopting 
the Cauchy approximation depend only on the value of $n$ (cardinality of the maximum) and can be quantified  calculating the 
relative error $\varepsilon(U_n)$ associated with the mode $U_n$ \cite{cooketal04_I}. In this case the approximated value is, 
from Eq.~(\ref{eq:penult2}), $\exp(-1)$ while the exact value is, from Eq.~(\ref{eq:penult1}), $(1-1/n)^n$. A plot of 
$\varepsilon(U_n)$ as a function of $n$ is reported in Fig. 1 of \cite{cooketal04_I}: e.g. for  $n=50$ the corresponding 
relative error is $\varepsilon(U_50)=0.01$, Note that for values $y>U_{n}$ the relative error is smaller than 
$\varepsilon(U_n)$ as $\Psi(y)<\Psi(u_{n})$.
 
We now consider the case of variates of exponential type: $\Psi(x)=\exp(-g(x))$ where $g(h)$ is a positive function 
which increases monotonically faster than $\ln(x)$. In this case from Eq.~(\ref{eq:penult2}) we obtain 
\begin{equation}
-\ln(-\ln(\zeta(y)))=h(y)-h(U_{n}).\label{eq:penult3}
\end{equation}
This last equation can be expanded in a Taylor series to obtain
\begin{equation}
-\ln(-\ln(\zeta(y)))=\frac{dh(y)}{dy}\bigg|_{U_{n}}(y-U_{n})+\frac{{d^{2}h(y)}}{dy^{2}}\bigg|_{U_{n}}
\frac{{(y-U_{n})^{2}}}{2!}+\dots.
\label{eq:penultimate_exponential_family}
\end{equation}
The extreme value theorem assures that for very large values of $n$, the linear term in 
Eq.(\ref{eq:penultimate_exponential_family}) dominates \cite{cooketal04_I,cramerbook}, therefore
\begin{equation}
\label{eq:ultimate_exponential_family}
-\ln(-\ln(\zeta(y))) \underset{n\rightarrow\infty}{=} \frac{dh(y)}{dy}\bigg|_{U_{n}}(y-U_{n})
\end{equation}
which is the ultimate approximation. Eq.~(\ref{eq:ultimate_exponential_family}) is the Gumbel distribution (EV1) with
location parameter $U_{n}$ and scale parameter $[(dh(y)/dy) |_{U_{n}}]^{-1}$: $U_{n}$ and $[(dh(y)/dy) |_{U_{n}}]^{-1}$ 
are the renormalization coefficients $b_{n}$ and $a_{n}$ of the extreme value theorem, Eq.~(\ref{eq:ev1}).
In the case of the Weibull distribution $h(x)=(x/C)^w$, thus using Eq.(\ref{eq:penultimate_exponential_family}) one 
obtains the following penultimate Taylor series approximation:
\begin{equation}
\label{eq:penult_weibull}
\begin{split}
-\ln(-\ln(\zeta(y)))= & \frac{{w(U_{n})^{w-1}}}{C^{w}}(y-U_{n})+\\ 
& \frac{{w(w-1)(U_{n})^{w-2}}}{C^{w}}\frac{{(y-U_{n})^{2}}}{2!}+\dots.
\end{split}
\end{equation}
For an exponential parent distribution ($w=1$) only the linear term of Eq.(\ref{eq:penult_weibull}) 
is non null (the derivative of $h(y)$ of order $>$1 are all null). In this case the penultimate and the ultimate 
approximation are equivalent. The convergence to the Gumbel distribution for the cumulative distribution 
of maxima extracted from an exponential parent is extremely fast: dictated by precision of the Cauchy 
approximation used to obtain Eq.(\ref{eq:penult2})). For values of the shape parameter $w\neq1$ the 
convergence to the Gumbel distribution is dictated by the rate with which the nonlinear terms in 
Eq.(\ref{eq:penult_weibull}) become negligible with respect to the linear term as $n\rightarrow \infty$. It is well 
know that this convergence might be very slow (even $n=10^{6}$ may not be sufficient for 
Eq.(\ref{eq:ultimate_exponential_family}) to be ``valid'') \cite{koutso04_I,koutso04_II,harris04,cooketal04_I}.

In this case, one should use the penultimate approximation of Eq.(\ref{eq:penult_weibull}) and not the GEV distribution 
of Eq.(\ref{eq:generalGEV}) as an accurate (neglecting the error due to the Cauchy approximation)  expression for the 
cumulative distribution of the $n$-sample maxima. However, if the shape parameter is not an integer value, then 
Eq.(\ref{eq:penult_weibull}) has an infinite number of terms and cannot be easily computed. To overcome this limitation 
we use the practice of '`preconditioning''  \cite{harris04,cooketal04_I}. We introduce the new variable $z=x^{w}$, whose 
distribution is exponential, 
$p(z)=\frac{{1}}{{C^{\prime}}^{-1}} exp(-z/C^{\prime})$, with $C^{\prime}=C^{w}$. 
For the variable $z$ the convergence of the cumulative distribution of n-sample maxima to the limiting Gumbel distribution
is very fast and we can write (using Eqs.(\ref{eq:mode}) and (\ref{eq:ultimate_exponential_family})
\begin{equation}
-\ln(-\ln(\zeta(y)))=\frac{{1}}{C^{\prime}}(y-C^{\prime}\ln n),
\label{eq:ultimate_perz}
\end{equation}
and finally
\begin{equation}
-\ln(-\ln(\zeta(y)))=\frac{{1}}{C^{w}}(y^{w}-C^{w}\ln n).
\label{eq:ultimate_figa}
\end{equation}
This equation is an exact, neglecting the error due the Cachy approximation, expression (when $n$ cannot be considered 
infinite, which is for all practical applications) for the probability $\zeta(y)$ for any value of the shape parameter $w$. 
Note that all the results obtained in this Section are also valid for varaites whose distribution function is right-tail 
equivalent to a Weibull \cite{cooketal04_I}. Two distribution functions $F_{1}$ and $F_{2}$ are right-tail equivalent if 
$(1-F_{1}(x))/(1-F_{2}(x))\rightarrow 1$ when $x\rightarrow +\infty$. The results presented in \cite{wiltuo05} 
indicate that the distribution function of the daily amount of precipitation is right-tail equivalent to the 
Weibull distribution function.

\subsection{Metastatistics}\label{sec:MEV}
The exceedance probability, $\zeta(y)$, of the n-maximum $Y_{n}$ depends on the cardinality, $n$, and on the parameters, 
$\overrightarrow{\theta}=(\theta_{1},..,\theta_{k})$, of the distribution of the parent variable $x$. To make this
dependence explicit we now adopt the notation $\zeta(y,n,\overrightarrow{\theta})$ instead of $\zeta(y)$. Let us consider, 
as an example, the series of maxima $\{Y_{j}\}=(Y_{1},Y_{2},...,Y_{T})$ each of them with variable cardinality 
$\{n_{j}\}=(n_{1},n_{2},...,n_{T})$ and whose parent variables, while sharing a common distribution, have different parameters
$\{\overrightarrow{\theta_{j}}\}=(\overrightarrow{\theta_{1}},\overrightarrow{\theta_{2}},...,\overrightarrow{\theta_{T}})$.
We want to find the probability $\hat{\zeta}(y)$ that none of the maxima in the sequence $\{Y_{j}\}$ has a value smaller 
than $y$. This example has practical relevance. In fact a typical hydrological application requires the estimate of daily 
rainfall amount with a return period of, say, 1,000 years from an observed time series of $T$ years. The number of wet days, 
days with a non-zero rainfall amount, changes from year to year, inducing a different cardinality of yearly maximum. 
Moreover inhomogeneity may be considered to be present, by which the distribution of daily rainfall amounts has a constant 
functional, but changing parameters every year. The classical results of Extreme Value Theory are not designed to handle this 
case, as they all postulate a constant cardinality of the n-sample maxima and to a homogeneous stochastic process.

Given the sequence of maxima $\{Y_{j}\}$ ($j=1,2,..,T$), the probability $\bar{\zeta}(y)$ for a maximum to not exceed the 
value $y$ is simply 
\begin{equation}
\label{eq:MEVpractical}
\bar{\zeta}(y)=\frac{{1}}{T}\sum_{j=1}^{T}\zeta(y,n_{j},\overrightarrow{\theta_{j}}).
\end{equation}
We use the term ``metastatistics factor'' to indicate $f(n,\vec{\theta})$ the probability density function of observing a 
maximum with cardinality $n$ and a parent distribution characterized by the parameters $\overrightarrow{\theta}$. With this 
definition we can write Eq.(\ref{eq:MEVpractical}) in the more general form 
\begin{equation}
\bar{\zeta}(y)=\iint dn\vec{d\theta}\; f(n,\overrightarrow{\theta})\;\zeta(y,n,\overrightarrow{\theta}),\label{eq:metastatistics}
\end{equation}
where the symbol $\overrightarrow{d\theta}$ denotes the differential $d\theta_{1}d\theta_{2}...d\theta_{k}$. Note that $n$ is 
integer variable but we keep for convenience a continuous notation with the understanding that the probability density function 
$f(n,\vec{\theta})$ is punctual in the variable $n$. We refer to Eq.(\ref{eq:metastatistics}) as the 
\textit{Metastatistics of Extreme Value (MEV)} formula. Note that Eq.(\ref{eq:metastatistics}) reduces to 
Eq.(\ref{eq:MEVpractical}) for the metastatistics factor 
$f(n,\vec{\theta})=\frac{{1}}{T}\sum_{j=1}^{T}\delta(n-n_{j},\vec{\theta}-\vec{\theta_{j}})$.

\subsubsection{Variable maximum order for Weibull variates with fixed scale and shape parameters}\label{sec:MEVapp1}
Hereby we consider a case where maximum values are extracted from a Weibull parent distribution with fixed scale $C$ and 
shape $w$ parameters but with a varaible order ($n$ not fixed). This case reflect a situation where the probability of the daily 
amounts of precipitation is homogeneous (invaraint from one year to the next) but the number of wet days (daily amount 
$>$0) in one yaer is not fixed (as it usually the case). Using Eqs.(\ref{eq:metastatistics}) and (\ref{eq:ultimate_figa}) 
we write
\begin{equation}
\label{eq:MEV_speccase1}
\begin{split}
\bar{\zeta}(y) = &\sum\limits_{n_{1}}^{n_{2}}f(n_{j})\exp(-\exp(-\frac{y^{w}}{C^{w}}+\ln n_{j})) = \\
&\sum\limits_{n_{1}}^{n_{2}}f(n_{j}) [\beta(y)]^{n_{j}}.
\end{split}
\end{equation}
In the above equation $\beta(y) = \exp(-\exp(-y^{w}/C^{w}))$, while $n_{1}$ and $n_{2}$ are the minimum and maximum 
values for the order $n$ (minimum and maximum number of wet days), and $f(n_{j})$ is the frequency with which the order 
$n_{j}$ is present in the maxima series. For $y \gg C$ the terms $[\beta(y)]^{n_{j}}$ can be approximated as follows
\begin{equation}
\label{eq:MEV_speccase2}
\begin{split}
[\beta(y)]^{n_{j}}= & [\exp(-\exp(-\frac{y^{w}}{C^{w}}))]^{n_{j}}\simeq[1-\exp(\frac{y^{w}}{C^{w}})]^{n_{j}} \simeq\\
& 1 - n_{j}\exp(\frac{y^{w}}{C^{w}}).
\end{split}
\end{equation}
When we insert this result in to Eq.(\ref{eq:MEV_speccase1}) we get 
\begin{equation}
\label{eq:MEV_speccase3}
\begin{split}
\bar{\zeta}(y)\simeq &\sum\limits_{n_{1}}^{n_{2}} f(n_{j}) \bigg[ 1 - n_{j}\exp(\frac{y^{w}}{C^{w}}) \bigg ]=1-
<n>\exp(\frac{y^{w}}{C^{w}}) \simeq\\
&\exp(-\exp(-\frac{1}{C^{w}}(y^{w}+C^{w}\ln <n>))),
\end{split}
\end{equation}
where $<n>$ indicates the average value of the maximum order. Notice that the last term of Eq.(\ref{eq:MEV_speccase3}) is 
identical to Eq.(\ref{eq:ultimate_figa}) with $<n>$ instead of $n$. Thus for $y\gg C$ the distribution function $\bar{\zeta}(y)$ 
of the mixture considered (fixed shape and scale parameters for the parent variable but varaible maximum order) is equal to 
the distribution function relative to a fixed order, this order being the average of the mixture of orders. The hypothesis 
that the parent variable is a Weibull variate has been essential in deriving this result. Therefore it may not be valid for 
variate which are not of the Weibull type. In the case where the shape $w$ and scale $C$ parameter in Eq.(\ref{eq:ultimate_figa}) 
are also variable, one can repeat the above arguments for $y\gg C_{max}$ ($C_{max}$ being the maximum scale in the mixture). 
If the order of maximum $n$ and the scale and shape parameters are independent from each other the metastatistics factor 
$f(n,C,w)$ can be consider as the product of two factors $f_{o}(n)$ and $f_{p}(C,w)$ and write
\begin{equation}
\label{eq:MEV_speccase4}
\bar{\zeta}(y)\simeq \int dCdw\,f_{p}(C,w)\exp(-\exp(-\frac{1}{C^{w}}(y^{w}+C^{w}\ln <n>))),
\end{equation}

\section{Results}\label{sec:results}
We first study the statistics of the daily amount of precipitation for our data set. Next, we show 
that the MEV formula, Eq.(\ref{eq:metastatistics}), together with Eq.(\ref{eq:ultimate_figa}) are 
the correct tool to estimate the distribution function of maxima drawn from a mixture of Weibull 
variates, while the GEV formula, Eq.(\ref{eq:generalGEV}), is not. Finally we apply the 
MEV approach to our data set and draw conclusions on its homogeneity and prediction on 
the structural stability for hydrological purposes. 

\subsection{Padova time series and Weibull approximation}\label{sec:padwei}
\begin{figure}[h]
\centering
\includegraphics[width=13cm]{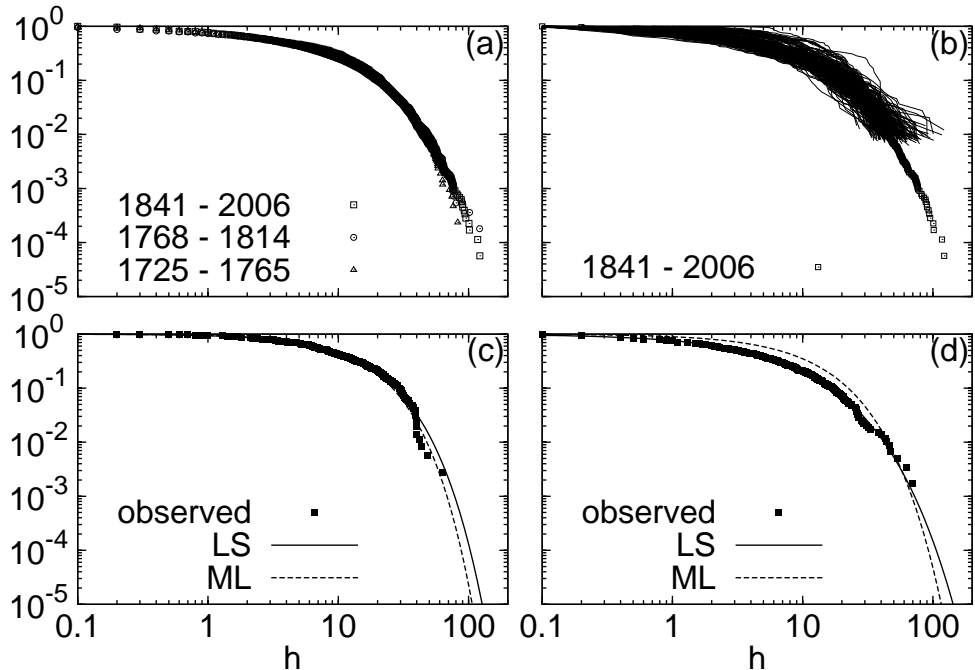}
\caption{The probability $\Psi(h)$ for the daily amount of rainfall to 
exceed a value $h$. Panel (a): $\Psi(h)$ for the three time intervals 
of observation considered for the Padova time series (points). 
Panel (b): $\Psi(h)$ for the entire 1841-2006 interval (squares) and for each year 
of the interval (solid lines). Panels (c) and (d): $\Psi(h)$ (full squares) together 
with the Weibull fits in the range $h>10$ mm by the least square method (LS) and by 
maximum likelihood method (ML).}
\label{figure1}
\end{figure}

The Padova time series has three intervals of continuous observation: 1725-1764, 1768-1814, and 1841-2006. For each interval 
we calculate the probability $\Psi(h)$ for the daily amount of rainfall to exceed a threshold $h$. The results are reported 
in panel (a) of Fig.\ref{figure1}. We see how all three intervals have similar complementary distribution functions. In 
panel (b) of the same figure we compare the probability $\Psi(h)$ relative to the 1841-2006 interval with the probabilities
calculated for each year of the interval: the ``cloud'' of yearly curves is approximatively symmetric with respect the curve 
relative to the entire interval. In panels (c) and (d) of Fig.\ref{figure1} we display the results of fitting the observed 
probability $\Psi(h)$ (squares) with a stretched exponential function ($\exp(h/C)^{w}$) for $h\geq$10 mm. Two fitting  
methodologies are adopted: the least square fit (solid line) and the maximum likelihood (dashed line). The result in 
panel (c) are relative to the 5 years interval 1841-1845, and those in panel (d) to 5 years interval 1926-1930. In most cases 
the least square fit and the maximum likelihood one produce similar results as in the case depicted by panel (c). However, due 
to the left truncation ($h\geq$10 mm) the algorithm (Matalb 2012) which minimize the likelihood is not always performing 
properly as in the case depicted in panel (d): the least square fit is a better approximation than the maximum likelihood fit. 
Moreover the algorithm maximizing the likelihood fails sometimes to find a maximum when too few data are available. E.g. when 
one considers only the data from a single year or two years, the condition $h>10$ mm  reduces the number of available points 
to $\sim$5-10. The least square fit, although it is not the most proper choice \cite{clau09}, does not 
suffer from these limitation and therefore we adopt in the following to show the validity of the metastatistics formula.

\begin{figure}[h]
\centering
\includegraphics[width=13cm]{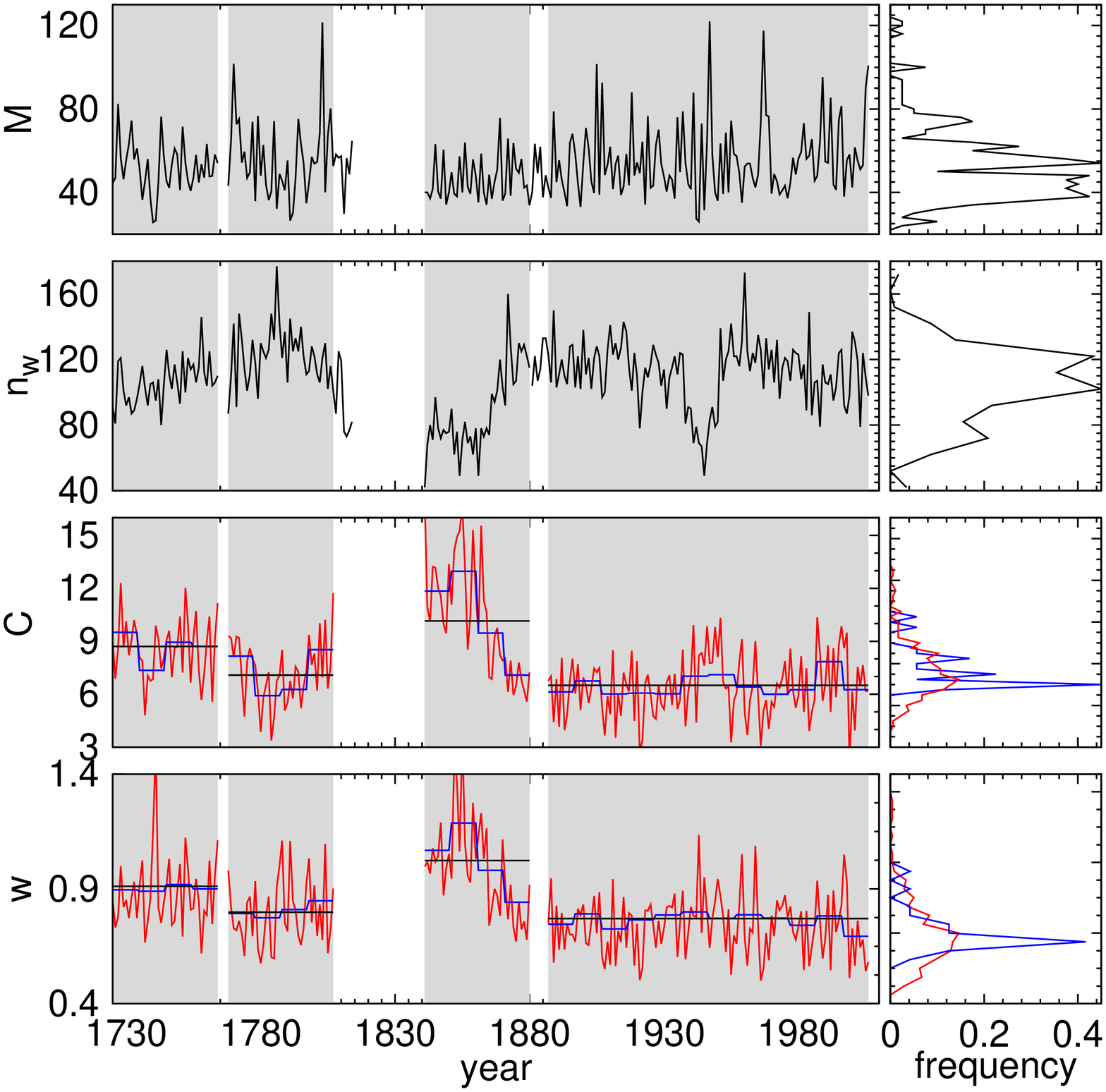}
\caption{The annual maxima $M$ (solid line in top left panel), the number of wet days $n_{w}$ per year, 
(solid line in middle-top left panel), the scale parameter $C$ (middle-bottom left panel), and the shape
$w$ parameter (bottom left panel) for the four intervals (indicated
by the shadowed region) of the Padova time series. For the shape and
scale parameter, we report the values obtained when using the distribution
relative to the entire data in the interval (horizontal black line),
the values obtained using 1 year windows (red solid line),
and the values obtained using 10 year windows (solid blue
line). Panels on the right report the observed frequencies for the
variables displayed on the left panels.}
\label{figure2}
\end{figure}

Next we consider the variability of the yearly maximum $M_{\textrm{yr}}$, number of wet days $n_{w}$ (days with a non null 
precipitated amount) together with that of the scale $C$ and shape $w$  parameter of Weibull fitted to the complementary 
distribution function $\Psi(h)$ given $h>10$ mm. The results are reported in Fig.~\ref{figure2}. In the left panels the areas 
shadowed in gray indicate four of the five subsets (see Sec. II) considered for separate analysis: 1725-1764, 1768-1807, 
1841-1880, and 1887-2006. Right  panels, report the observed frequencies corresponding to the quantities depicted in 
left panels. The top left panel depicts the variability of the annual maxima. The distribution has a plateau in the region 
40-65 mm with a positive skewness. The middle-top panel  depicts the variability of $n_{w}$. The mode 
is $\sim$100 days with a second peak of almost equal intensity at $\sim$120 day. 

To address  the issue of homogeneity for the daily amount of rainfall dynamics we operate as follows. We first consider the 
value of the scale $C$ and shape $w$ parameters for each of the five subsets considered for our analysis (Sec.~\ref{sec:data}). 
Then, each subset is divided in 10,5,2, and 1 year long not overlapping windows. Inside each window the scale and 
shape parameters are calculated. The results of this procedure are reported on the midlle-bottom panel (c), scale, and bottom 
panel (d), shape, of Fig.~\ref{figure2}. The blue line refers to the result obtained with 10 years moving window, the red line 
to those obtained with 1 year moving window, and the black line to the result for the entire subset. For a better visualization 
the results obtained with 2 years and 5 years moving windows are not reported. From a visual 
inspection of these results we can formulate the following hypotheses. The intervals 1725-1764, 1768-1807, and 1887-2006 are 
intervals during which the daily amount of precipitation can be considered homogeneous: the variability 
of the scale and shape parameters seems to be quite symmetrical with respect the values calculated using the entire interval. 
This hypothesis is maybe true also for 1841-1880 interval, while the interval 1841-1920 is one for which the 
rainfall process at a daily scale cannot be considered homogeneous (even if the results relative to this interval are 
explicitly reported the observations relative to the 1841-1880 interval and the first 40 year of the 1887-2006 suggest 
this conclusion). In Section \ref{sec:homononhomo} we will test more rigorously these hypotheses. 
\subsection{MEV distribution vs GEV distribution}\label{sec:MEVvsGEV}
In the following we will compare the metastatistics approach, Eq. (\ref{eq:metastatistics}) with the one based on fitting the 
maxima series to the generalized extreme value distribution, Eq. (\ref{eq:generalGEV}). We show that the metastatistics is the 
correct approach in case of inhomogeneity. In particular we show that, in the case of maxima extracted from a mixture of parent 
Weibull variates, the adoption of the penultimate approximation, Eq.(\ref{eq:ultimate_figa}) coupled with the metastatistics 
formulation, Eq.(\ref{eq:metastatistics}), are the proper tools to address the question of the projected frequency of extreme 
events.

For this purpose we consider three experiments using artificially generated sequences. Experiment (1) Maxima are extracted with a 
fixed cardinality and from a Weibull parent variable with fixed scale and shape parameter. This experiment correspond to consider 
a homogeneous rain dynamics with a fixed amount of wet days for each year. Experiment (2) Maxima extracted with a fixed 
cardinality and from a Weibull parent variable which scale and shape parameter changing every 5 maxima extractions. This 
experiment corresponds to consider a rain dynamics which homogeneous (stable) for 5 years after which a new condition is 
achieved. The number of wet days is fixed. Experiment (3) Maxima extracted with a fixed cardinality and from a Weibull parent 
variable which scale and shape parameter changing every 2 maxima extractions. This experiment is analogous to the previous 
one except that the rain dynamics is stable only for 2 years. To mimic conditions which are typically encountered in rainfall 
time series, we set the number of maxima to be 50 (50 years of data) and the cardinality of the maxima to be 100 (100 wet 
days per year). The scale and shape parameters are those obtained from the Padova time series adopting the 50 years interval 
from xxxx to yyyy. Then we proceed as follows. For each experiment we generate the corresponding sequence 
of 50 ``years'' each with 100 ``days'' of non null precipitation. These variates are used to calculate the scale $C$ and 
shape $w$ parameters which  are fed in Eq. (\ref{eq:ultimate_figa}) and into Eq.(\ref{eq:MEVpractical}) to calculate the 
MEV estimate of the distribution funciton $\bar{\zeta}(y)$, the cumulative distribution of maxima. Moreover for each year, 
we calculate the maximum to obtain a sequence of maxima which we fit (using the minimum likelihood method) to the generalized 
extreme value distribution (Eq. (\ref{eq:generalGEV})) and to the Gumbel distribution to obtain the GEV and Gumbel estimates 
of $\bar{\zeta}(y)$ respectively. We repeat this procedure 1,000 times so that for each value $y$ we can calculate the median 
value of $\bar{\zeta}(y)$ of the 1,000 realizations. The median values relative to the MEV, GEV, and Gumbel methodology are 
compared in a Gumbel plot: $-\ln(-\ln(\bar{\zeta}(y)))$ versus $y$. To assess which of these three methodologies is the most 
accurate we generate, for each experiment, a sequence $10^7$ maxima (this is done simply buy parsing together series 
of 50 maxima generated according the prescription of each experiment) $m_{j}$ which we sort in ascending order. The sorted 
sequence is used to create the couples of points ($-\ln(-\ln((j-0.5/10^{7}))),m_{j}$) which are used as "truth" in the 
Gumbel plot of $-\ln(-\ln(\bar{\zeta}(y)))$ versus $y$.

\begin{figure}[h]
\centering
\includegraphics[width=13cm]{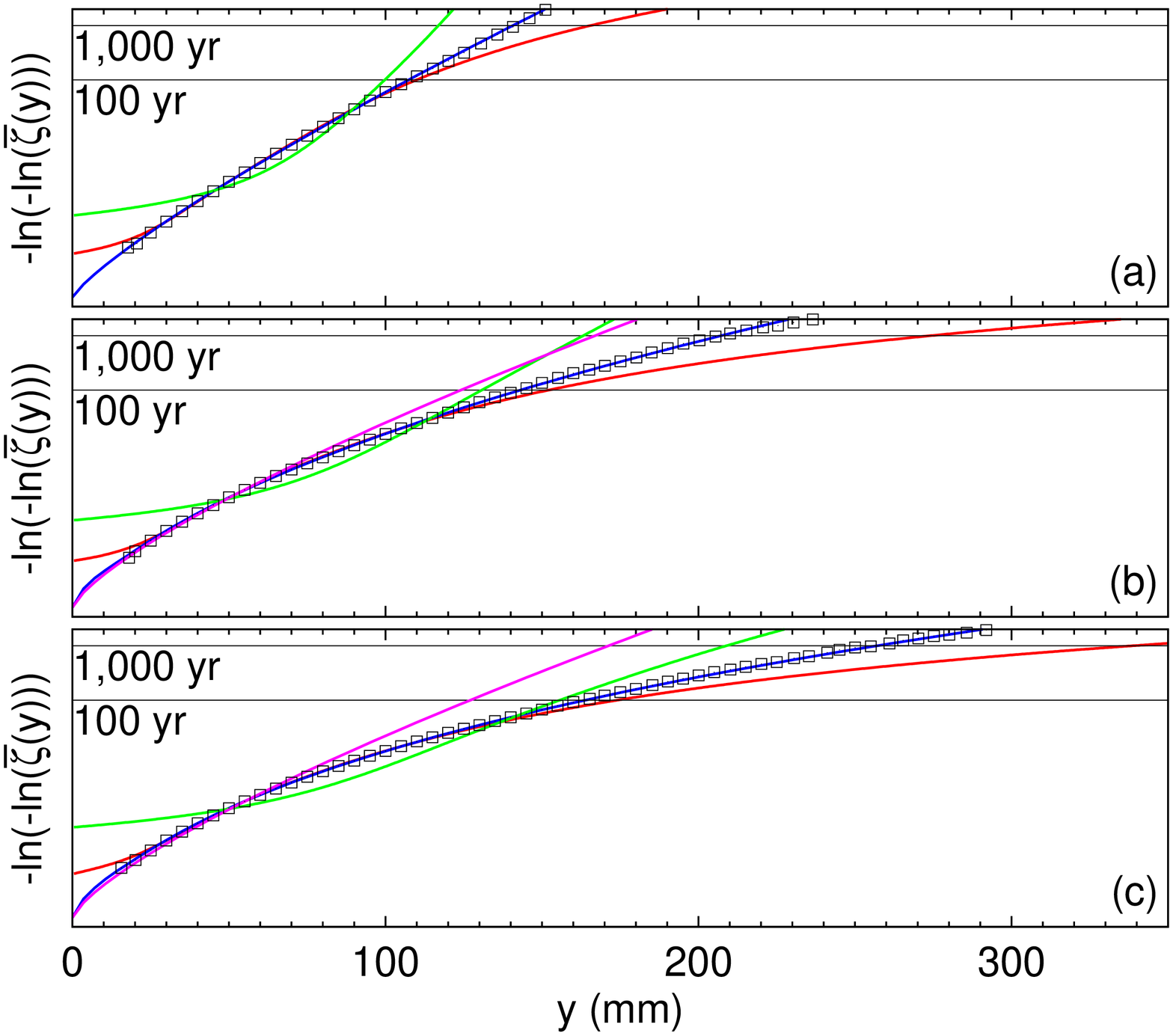}
\caption{Gumbel plot of the probability $\bar{\zeta}(y)$ of exceeding the value
$y$ for a maxima for three different artificially generated sequences. Panels (a), 
(b), and (c) refer respectively to the experiment (1), (2), and (3). The solid lines indicate: 
the median of the preconditioned metastatistics estimate (blue), the median of 
the generalized extreme value estimate (red), the median of the Gumbel estimate (green), 
the median of the preconditioned metastatistics estimate obtained using average values 
of scale and shape parameter (pink). Black empty square denote the observed expected results (truth).}
\label{figure3}
\end{figure}

The results are show in Fig.~\ref{figure3}. Panels (a), (b) and (c) refer respectively to experiments (1), (2) and (3).
Blue curves indicate the MEV median estimate, red curves the GEV median estimate, and green curves the Gumbel median estimate. 
Black squares represent the "truth" values. Pink curves in panels (b) and (c) are MEV median estimate obtained using average 
values of the scale and shape parameters. We see how in all case the MEV median estimate coincides with the expected values 
(truth). The GEV median estimates consistently underestimates the correct probability $\bar{\zeta}(y)$ (overestimates the 
precipitation value associated with a given return period), while the Gumbel median estimate consistently overestimates it 
(underestimate the precipitation value associated with a given return period). 

The results of Sec.~\ref{sec:MEVapp1} show that for the artificial sequences adopted in the experiments (1), (2), and (3) the 
influence of the a varaible cardinality amount to the adoption (for $y \gg C_{max}$, $C_{max}$ being the maximum value of 
the scale parameter in the mixture) of  the penultimate approximation formula, Eq.(\ref{eq:ultimate_figa}) with a cardinality 
equal to the average cardinality of the maxima sample. We verified this prediction running experiments (1)--(3) with a variable 
cardinality. The results are not reported for brevity.

\subsection{The question of homogeneity}\label{sec:homononhomo}
In the previous Section we have demonstrated the superiority of the metastatistics approach 
over the generalized extreme value and Gumbel prescriptions. We are now in a position to address the question 
of homogeneity in the Padova time series. We selected 5 intervals: 1725-1764, 1768-1807, 1841-1880, 1841-1920, 
and 1887-2006. With the help of Fig.~\ref{figure2}, we formulated the following hypotheses. During the intervals 
1725-1764, 1768-1807 and 1887-2006 the sequence of daily amount of rain appears to be homogeneous, during the 
interval 1841-1920 appears to be inhomogeneous, while we are undecided regarding the the 1841-1880 interval. 

To check the validity of these hypotheses, we calculate for each of the five intervals the scale $C$ and shape parameter 
$w$ of the stretched exponential function fitting the probability $\Psi(h)$ for the daily amount of rain to exceed a 
threshold $h$, given $h>10$ mm. This is equivalent to consider, at least initially, the daily amount of rain in each 
interval as a homogeneous process. Then given an interval, we use the computed  scale and shape parameters to generate 
1,000 artificial homogeneous sequences. Each of these sequences is then divided in non overlapping subsets of duration 
10,5,2, and 1 year. For each subset length we consider the MEV estimate of the probability $\bar{\zeta}(y)$ (the probability 
for a maximum to not exceed the threshold $y$) is calculated via Eqs. (\ref{eq:ultimate_figa}) and (\ref{eq:MEVpractical}). 
The results of the 1,000 repetitions are used to calculate the 5\%, 50\%, and 95\% percentile of $\bar{\zeta}(y)$ given $y$. 
The rationale is that if inside an interval  the daily amount of rain is an homogeneous process we expect the MEV 
estimate of $\bar{\zeta}(y)$ calculated with subsets of $m$ (10,5,2, and 1) years to be inside the 5\%, 95\% percentile 
range of the MEV estimate calculated under the hypothesis of homogeneity (note that since we use the penultimate 
approximation in the MEV formula, a MEV estimate in homogeneous condition is different from a GEV estimate).

\begin{figure}[h]
\centering
\includegraphics[width=13cm]{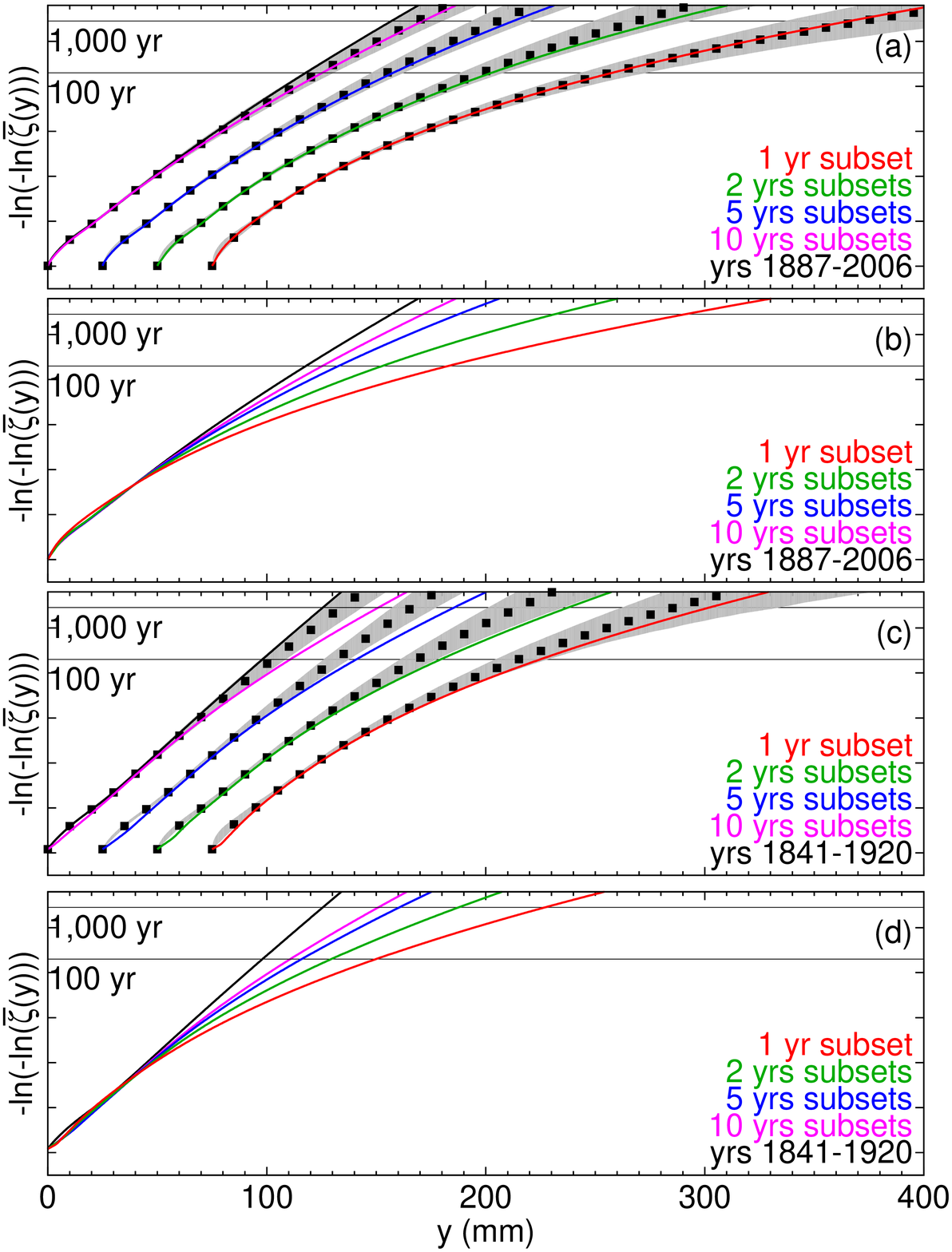}
\caption{Gumbel plots of the probability $\bar{\zeta}(y)$ of exceeding the value $y$. Panel (a):  preconditioned 
metastatistics estimate for the time interval 1887-2006  using the whole set (black line), 10 years long subsets 
(pink line), 5 years long subsets (blue line), 2 years long subsets (green line), and 1 year long subsets (red line). 
The areas shadowed in gray indicate the spreading (5\% to 95\% percentile), expected in the homogeneous case, while 
the black squares indicate the median. The curve relative to 5, 2, and 1 years subset are shifted for clarity. Panel 
(b): as panel (a) but only preconditioned metastatistics estimates with no shift. Panels (c) and (d): as panels (a) 
and (b) but for the 1841-1920 time interval.}     
\label{figure4}
\end{figure}

Figure \ref{figure4} presents the results of this analysis adopting a Gumbel plot: $-\ln(-\ln(\bar{\zeta}(y)))$ versus $y$. 
In panels (a) and (c), the gray shadowed area depict the 5\%, 95\% percentile range calculated via montecarlo simulation under 
the hypothesis of homogeneity, while the black square indicate the 50\% percentile (median). Moreover the colored solid lines 
represent the MEV estimate done with subsets of different lengths of the time interval considered. (red for 1 year, green for 
2 years, blue for 5 years, pink for 10 years, and black for the entire interval). Finally the curves relative to 5, 2, and 1 
years subsets are shifted by  25, 50, 75 mm for clarity. From panel (a) we conclude that for the interval 1887-2006  the 
daily amount of rain can be considered a homogeneous stochastic process. In fact for all different subset lengths the MEV 
estimate reside inside the 5\%, 95\% percentile interval of the homogeneous hypothesis. Similar results (not reported here 
for brevity) hold for the intervals 1725-1764, 1768-1807, and 1841-1880. If we consider the interval 1840-1920, panel (c), 
we see that the hypothesis of homogeneity is not tenable as the MEV estimate relative to 10 years and 5 years long subsets 
(pink and blue curves) resides clearly outside the 5\%, 95 \% percentile range expected  for the homogeneous case. The estimate 
relative to 2 years long subsets (green curve) is at the border  of the 5\%, 95 \% percentile range, while the one for 1 year 
long subset (red) is inside (we will offer an explanation for this effect in the following). 

Finally, panel (b) and (d) report only the median value of the estimates relative to different subset lengths without any 
shift. As the length of the subset, used for the  preconditioned metastatistics methodology, decreases (10 years to 1 year) 
the estimate median probability $\bar{\zeta}(y)$ decreases in value for any fixed $y$ (the daily amounts of rain relative 
to a given return period increase). This effect is due to progressive decrease of statistical accuracy, moving from 10 
years to 1 year, in fitting the tails of the probability $\Psi(h)$ for the daily rainfall amount to exceed a threshold 
$h$ given $h>$ 10 mm. As the length of the subset decreases larger fluctuations (with respect the value relative to the 
entire set, see also panel (b) and (c) of Fig.~(\ref{figure2}) for the shape and scale parameter are observed. These 
fluctuations are approximatively symmetric, however their effects on the annual maxima are not. The right tail of 
$\bar{zeta}(y)$ is dominated by those fluctuations of the scale and shape parameters increasing the probability of 
observing larger (with respect to the entire set case) maxima. In turn, this implies a decrease for the cumulative 
distribution $\bar{\zeta}(y)$. These last results might explain why, for the interval 1840-1920, the 2 years and 1 year 
long subsets estimate are respectively close to, and inside the 5\%, 95\% percentile range expected for a homogeneous 
process: the fluctuation due to the lack of statistics might be large enough to hide the inhomogeneity of the data. 

\subsection{Predictions for structural stability}\label{sec:structsta}
In the previous Section we have shown that for the time intervals 1725-1764, 1768-1807, 1841-1880, and 1887-2006 the 
daily amount of rainfall can be considered a homogeneous process: fixed scale and shape parameter. For these intervals, 
the probability $\bar{\zeta}(y)$ for the yearly maximum daily amount to not exceed the threshold $y$ can be calculated 
with the MEV formalism as 
\begin{equation}
\begin{split}
\bar{\zeta}(y)=&\frac{{1}}{T}\sum_{j=1}^{T}\zeta(y,n_{j},C_{j},w_{j})=\frac{{1}}{T}\sum_{j=1}^{T}\zeta(h,n_{j},C,w) \\
& \frac{{1}}{T}\sum_{j=1}^{T} \exp(-\exp(-\frac{1}{C^{w}}(y^{w}-C^{w}\ln n_{j}) \simeq\\
& \exp(-\exp(-\frac{1}{C^{w}}(y^{w}-C^{w}\ln <n>) 
\end{split}
\label{eq:homometa}
\end{equation}
For the time interval 1840-1920 the daily amount of rain could not be considered a homogeneous process. The analysis 
of the previous Section and Fig.~\ref{figure4} suggest that, for this time interval, the MEV estimate calculated with 
5 years long subsets should be used. The results relative to the 1 year and 2 year subsets are too noisy to be trustworthy. 
The results relative to 10 years long subsets are appreciably different from those relative to 5 years: the 5 year long 
subset better resolve the inhomogeneity of the interval.

\begin{figure}[h]
\centering
\includegraphics[width=13cm]{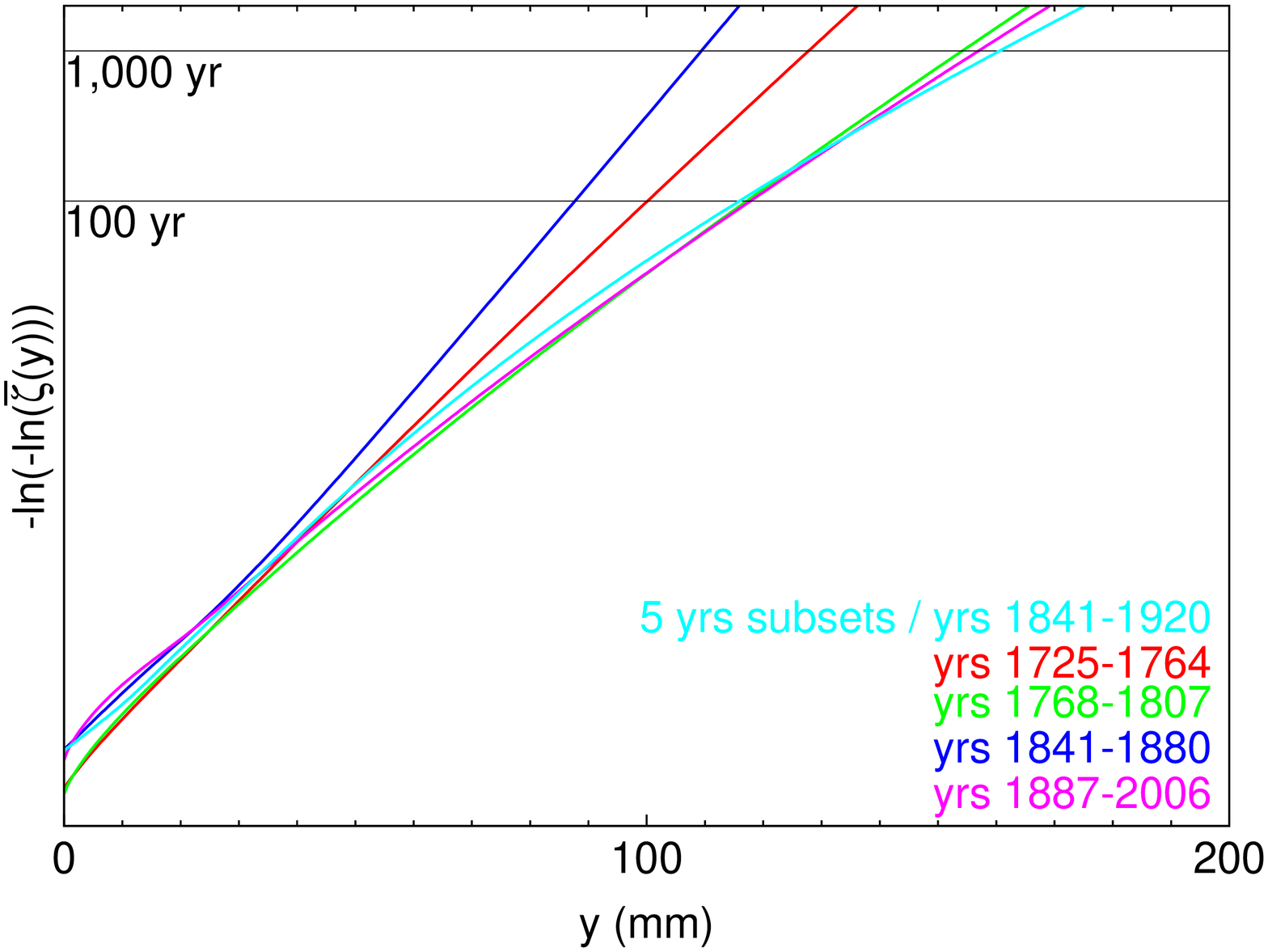}
\caption{Gumbel plot of the probability $\bar{\zeta}(y)$ of exceeding the value $y$ for the time intervals 1725-1761, 
1768-1807, 1841-1880, 1841-1920, and 1887-2006.}
\label{figure5}
\end{figure}

Figure \ref{figure5} depicts the Gumbel plot for the probability $\bar{zeta}(y)$ for the time 
intervals 1725-1764 (red curve), 1768-1807 (green curve), 1841-1880 (blue curve) 1887-2006 (pink curve), together with the 
5 years long subset estimate for the interval 1841-1920 (cyan curve). We first consider the 1,000 years return time 
intensity. These intesities are all estimates from MEV fits. The most conservative prediction is the one for the interval 
1841-1880 ($\sim$110 mm), followed by the 1725-1764 ($\sim$130 mm). The estimates for the remaining intervals predict an 
intensity of $\sim$ 160 mm. Overall there is a $\sim$50 mm variability, not exactly a negligible one. In the cas of the 
100 year return time intesity the varaibilty is $\sim$30 mm (from 90 mm to 120 mm) which is also not negligible. Note that 
except for the 1887-2006 interval, which is 120 years long, all the 100 years return time intensites are estimates 
from the MEV fit. Overall, the variability observed is not connected to the procedure adopted to derive the predictions 
(preconditioning and metastatistics) which is the proper one, but to the fact that we sample different epochs with different 
climate ``conditions''. Even when we use the MEV estimates, we assume that the \textit{future} will 
be the same as the \textit{present} (the time interval used to make the prediction) but this is rarely the case. 

\section{Conclusions}\label{sec:conclu}
The metastatistics approach described in this manuscript extend the extreme value theorem 
to statistical inhomogeneous cases. These are the most probable cases occurring in nature. In particular, 
we have applied the metastatistics to the case of Weibull variates . In this case the metastatistic approach  
coupled with the practice of preconditioning offers the correct solution while the standard method 
(fitting the maxima with the generalized extreme value distribution) adopted in literature does not.
The case of Weibull variates is of particular importance because the distribution of daily amount of rainfall 
is right tail equivalent to a Weibull distribution \cite{wiltuo05}. Thus the metastatistics approach together with the 
penultimate approxiamtion are the proper tools to address the important question of predicting the frequency of extreme 
hydrological events. We have done so using the Padova time series. Five different predictions have been derived: one 
for each time interval of time series considered. The variability observed for the intensity of the 
1,000 (100) year return time event is of the order of 50 (30) mm: not a negligible one.

These limitations reflects the fact that different climate condition have been adopted, and that  
one (as in all works in literature) consider the climate condition with the prediction is done to 
be valid also in the future. Using the 1841-1880 time interval we have 110 mm daily amount 
with a return time of 1,000 years. But, using the 1887-2006 time interval we would predict 160 mm 
as the daily amount with occurring on average once in a millennium. How to bypass this limitations 
in the case of hydrological extreme? We need to connect the daily precipitation dynamics to the 
climate parameters which can more accurately be estimated by climate models. 
In practice the shape $C$ and scale $w$ parameter are dependent on some of the climatological parameters 
$\overrightarrow{\pi}$. Note that any dependece is likely to be stocahstic in nature rather than a deterministic 
one. Thus in theory we can use the dependance and the ``proper'' estimate of the future value of the climatological 
parameters to ``estimate'' what would be the future metastatistics factor $f_{\textrm{fut}}(C,w)$ to use in the MEV 
formula, Eq. (\ref{eq:metastatistics}). This will enable one to make a prediction of  the frequency 
of extreme events which matches the ``future'' climate condition and not the current one. 














\end{document}